\documentclass[]{elsart}

\usepackage{epsfig}
\usepackage{graphics}
\usepackage{graphicx}
\usepackage{amsmath}
\usepackage{url}
\usepackage{subfigure}

\pdfoutput=1

\begin{document}

\begin{frontmatter}

\title{Game of Life on Phyllosilicates:  Gliders, Oscillators and Still Life}

\author{ Andrew  Adamatzky}
\address{University of the West of England, Bristol, UK}

\maketitle

\begin{abstract}

\noindent
A phyllosilicate is a sheet of silicate tetrahedra bound by basal oxygens. A  phyllosilicate 
automaton  is a regular network of finite state machines --- silicon nodes and oxygen nodes --- which mimics structure of the phyllosilicate. A node takes states 0 and 1. Each node 
updates its state  in discrete time  depending on a sum of states of its three (silicon) or 
six (oxygen)   neighbours. Phyllosilicate automata exhibit localizations attributed to Conway's Game of Life: gliders, oscillators, still lifes, and a glider gun. Configurations and behaviour of typical localizations, and interactions between the localizations 
are illustrated.

\vspace{0.5cm}

\end{abstract}

\end{frontmatter}

\section*{Adamatzky A. Game of Life on Phyllosilicates:  Gliders, Oscillators and Still Life. Physics Letters A 377 (2013) 1597--1605.}

\section{Introduction}

Phyllosilicates are parallel sheets of silicate tetrahedra, they are widely present in nature and 
typically found in clay-related minerals on the Earth surface \cite{griffen_1993, Bergaya_2006, Bleam_1993, richardson_2008}.
Phyllosilicates are used in development of nano-materials, nano-wires and patterned surfaces for 
nano-biological interfaces~\cite{Monnier_1993, Carrado,Suh_2009}. They are also employed, in a form 
of cation-exchanged sheet silicates, as catalysts in chemical reactions~\cite{Ballantine_1984}, e.g. 
nickel phyllosilicate catalysts~\cite{Lehmann_2012,McDonald_2009,Specht_2010}. 

Silicate lattice exhibit a range of  defects, or localised imperfections. They are  lattice vacancies, or vacancy-impurity pairs, interstitial defect, and substitutional defects~\cite{sokolski_1967, reiche_2011}. Also recombination induced defects might form when a strong electrical pulse is applied to the lattice~\cite{tanimura_1983}. 
The defects can migrate through the lattice, especially in presence of  ionising radiation~\cite{watkins_1976, watkins_1977}, and interact with other defects, sometimes forming second-generation defects ~\cite{sokolski_1967}; see also ~\cite{reiche_2011}. In~\cite{adamatzky_2013_IJMP, adamatzky_IJBC_2013}  we introduced phyllosilicate automata and studied their excitable, threshold and intervals of 
excitation~\cite{adamatzky_2013_IJMP}, and binary-state~\cite{adamatzky_IJBC_2013}.  By sampling of thousands of rules of binary phyllosilicate automata~\cite{adamatzky_IJBC_2013} we discovered rules that show (1)~sub-linear growth of disturbances, and (2)~exhibit stationary and (3)~travelling localizations. As far state-transition rules of phyllosilicate binary automata satisfy conditions (1--3), they an be classed non-rectangular lattice analogs of Conway's Game of Life cellular automata~\cite{gardner_1970}. We continue lines of enquiry into space-time dynamics of cellular automata on non-orthogonal and aperiodic lattices, including triangular tessellations and Penrose tilings~\cite{bays_2007, owens_2010, goucher_2012}. We introduce and study an automaton network --- phyllosilicate automata ---  where connections between finite automata are inspired by simplified structure of silicate sheets, lattices of connected tetrahedra, and investigate the space-time dynamics of compact clusters of non-resting nodes on these automaton networks. 

The paper is structured as follows. In Sect.~\ref{model} we define phyllosilicate automata and Life-like rules.
Three types of gliders are detailed in Sect.~\ref{gliders}. Section~\ref{oscillators} introduces stationary oscillating localizations. Stationary still localizations are studied in Sect.~\ref{stilllife}.

\section{Phillosilicate automata}
\label{model}

\begin{figure}[!tbp]
\centering
\includegraphics[width=0.5\textwidth]{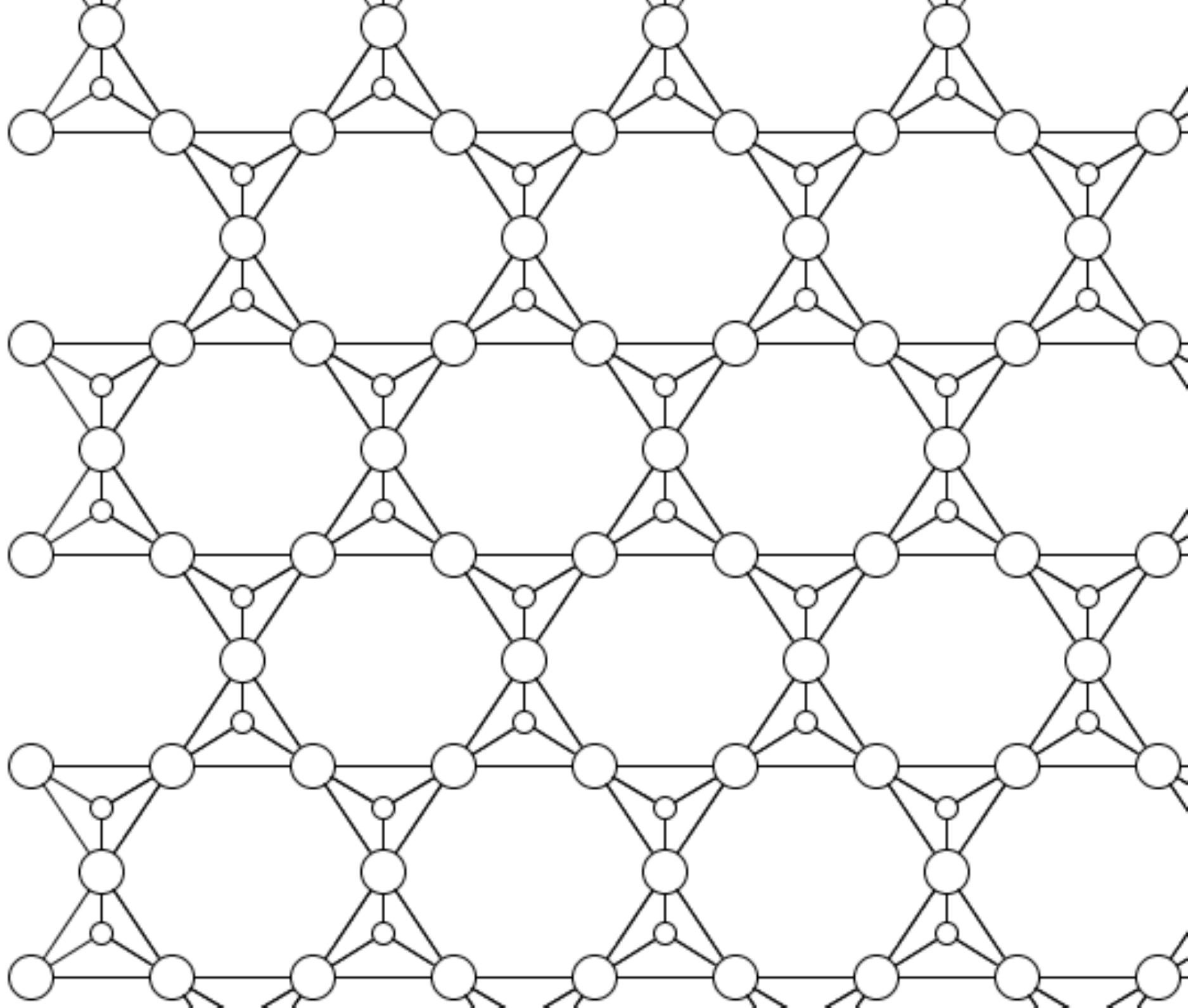}
\caption{Phillosilicate automata. Automata representing silicon are small discs, 
oxygen automata are large discs.  Unshared oxygen atoms of the tetrahedra are not shown.}
\label{latticestructure}
\end{figure}

Phyllosilicates are sheets of coordinated  (SiO$_4$)$^{4-}$ tetrahedra units. Each tetrahedron shares its three corner basal oxygens of neighbouring tetrahedra~\cite{pauling_1930, Liebau_1985, Bleam_1993} (Fig.~\ref{latticestructure}).  Vacant oxygens of all tetrahedra point outside the lattice, away of tetrahedra; these apical oxygens are not taken into account in our automaton models.  

A phyllosilicate automaton $\mathcal A$ is a two-dimensional regular network (Fig.~\ref{latticestructure}) finite state machines, automata; we call them nodes. There are two types of nodes in the network: silicon nodes $s$ (centre vertex of each tetrahedron)  and oxygen nodes  $o$ (corner vertices of each tetrahedron). A silicon node has three neighbours. An oxygen node has six neighbours (Fig.~\ref{latticestructure}).

The nodes take two states   0 (resting state) and 1 (non-resting state) and update their states $s^t$ and $o^t$ simultaneously and in discrete time $t$ depending on states of their neighbours. Let $\sigma^t(a)$ be a number of 1-state neighbours of node $a$, $0 \leq \sigma^t(s) \leq 3$ and  $0 \leq \sigma^t(o) \leq 6$. Let $o$ and $s$ be instances of oxygen and silicon nodes, and $\mathbf{o}$ and $\mathbf{s}$ be binary strings of seven and four symbols each; $\mathbf{a}[i]$ is a symbol at $i$th position of string $\mathbf{a}$. A node $a$, $a=o$ or $s$, updates its state by the rule: 
\begin{equation}
a^{t+1}=\mathbf{a}_{a^t}[\sigma(a)^t] ,
\label{rule}
\end{equation}
where $a = s$ or $o$ and $\mathbf{a} = \mathbf{s}$ or $\mathbf{o}$. When referring to any particular rule 
$(\mathbf{s}_0, \mathbf{s}_1, \mathbf{o}_0, \mathbf{o}_1)$, where string $\mathbf{a}_i$ determines next state of automaton being in state $i$,  we will be using decimal representations of strings as 
$$\mathcal{R}(dec(\mathbf{s}_0), dec(\mathbf{s}_1), dec(\mathbf{o}_0), dec(\mathbf{s}_1))$$.  We assume state 0 is a quiescent, or resting, state: a node in state 0 will not take state 1 if all its neighbours are 0. We will be calling nodes in state  0 resting nodes, and nodes in state 1 non-resting nodes. 

\begin{table}[!tbp]
\caption{Parameters of localizations. For each localisation $L$ we indicate its period $p$, minimum $w_{\min}$ and 
maximum $w_{\max}$ numbers of non-resting nodes, minimum $w_{\min}^o$ and 
maximum $w_{\max}^o$ numbers of non-resting oxygen nodes, and minimum $w_{\min}^s$ and 
maximum $w_{\max}^s$ numbers of non-resting oxygen nodes. Minimal values in $\min$ columns and 
maximum values in $\max$ columns are underlined.}
\begin{tabular}{l|lllllll}
$L$ 			& $p$   &  $w_{\min}$ &  $w_{\max}$  &$w_{\min}^o$  &$w_{\max}^o$  & $w_{\min}^s$ & $w_{\max}^s$  \\ \hline \hline
 $G_{65}^{16}$ 	&  \underline{16} &  6 &  13 &  2 & 9 & 2 & 8 \\ 
  $G_{68}^{12}$ 	&  12 & 7 & 13 & 2 & 6 & 2 & 7 \\
 $G_{72}^4$ 		& 4 & 4 & 11 & 2 & 6 & 2 & 6\\
  $O_{65}^5$ 	& 5 & 6   & 10  & 2 & 8 & 4& 8\\ 
  $O_{65}^6$ 	&  6 & 4  & 9  & 2  & 5 & 2 & 4 \\
  $O_{65}^{6A}$  	& 6 &  3 & 5 & \underline{1} & 3 & 2 & 2 \\  
 $O_{68}^6$		& 6 & 3 & 6 &  \underline{1} & 3 & 2 & 3 \\
 $O_{68}^{12}$	& 12  & 8  & \underline{26}  & 4  & \underline{22} & 0 & \underline{16} \\
 $O_{72}^2$		&  2 & 3 & 4 & 3 & 3 & 4 & 4 \\
 $O_{72}^4$		& 4 & 6  & 10 & 3 & 6  & 4 & 9\\
 $O_{72}^{12}$	& 12 & \underline{2} & 5 &  \underline{1} & 3 & \underline{1} & 3 \\         
\end{tabular}
\label{table}
\end{table}

In our previous paper~\cite{adamatzky_IJBC_2013} we studied over 100K rules, randomly sampled,   of binary-state phyllosilicate automata.  We uncovered principle morphologies of patterns emerging and  classified rules based on shape of patterns generated by the rules and internal morphologies of the patterns, and approximated distribution of functions on density of patterns and activity of cell-state transitions and maps of oxygen and silicon densities. Using shapes of patterns as a key characteristic we selected five classes: octagonally shaped patterns, almost circularly shaped, stationary patterns non-propagating beyond convex or concave hulls, patterns exhibiting dendritic growth,  and patterns showing still localizations and oscillations. Internal morphology driven taxonomy is comprised of solid patterns, labyrinthine patterns, wave-like patterns, disordered patters; and still, travelling and propagating localizations. 

The 100K were mapped onto activity versus density space~\cite{adamatzky_IJBC_2013}, where activity reflects a number of nodes changing
their states each time step and density is a number of non-quiescent nodes in a fixed-size domain. Exploring the 'activity-density' space in the loci
with high activity and low density we discovered three minimal --- minimality in number of '1's in strings $\mathbf{s}$ and $\mathbf{o}$ --- rules that support gliders, $\mathcal{R}_{65}=\mathcal{R}(5, 2, 16, 65)$, $\mathcal{R}_{68}=\mathcal{R}(5, 2, 16, 68)$, and $\mathcal{R}_{72}=\mathcal{R}(5, 2, 16, 72)$.  The rules have the same rules for state transitions of silicon nodes, $\mathbf{s}_0 = (0101)$ and $\mathbf{s}_1 = (0010)$,  and oxygen nodes in resting states, $\mathbf{o}_0 = (0010000)$. Only  behaviour of oxygen nodes in   non-resting state differs: $\mathbf{o}_1 =(1000001)$ in rule $\mathcal{R}_{65}$, $(1000100)$ in $\mathcal{R}_{68}$,  and $(1001000)$ in  $\mathcal{R}_{72}$. Each of thee three rules  
$\mathcal{R}_{65}$,  $\mathcal{R}_{68}$ and $\mathcal{R}_{72}$. generate its own unique glider and several oscillators. Parameters of gliders and oscillators are shown in Tab.~\ref{table}. We have also discovered one still life, observed in rule $\mathcal{R}_{68}$  and a glider gun, found in rule $\mathcal{R}_{72}$. 

Gliders, oscillators and still lifes are characteristic patterns of Conway's Game of Life cellular automata. Phyllosilicate analogs of the Game of Life, or Life-like automata, obey the following rules: 
\begin{itemize}
\item Silicon node in state 0 takes state 1 only if it has one or three neighbours in state 1. 
\item Silicon node in state 1 remains in state 1 if it has two neighbours in state 1. 
\item Oxygen node in state 0 takes state 1 only if it has two neighbours in state 1.
\item An oxygen node in state 1 remains in state 1 if it has no neighbours in state 1
or six in rule $\mathcal{R}_{65}$, four in rule $\mathcal{R}_{68}$ and three in rule $\mathcal{R}_{72}$ neighbours
are in state 1. 
\end{itemize}

\section{Gliders}
\label{gliders}

\begin{figure}[!tbp]
\centering
\includegraphics[scale=1]{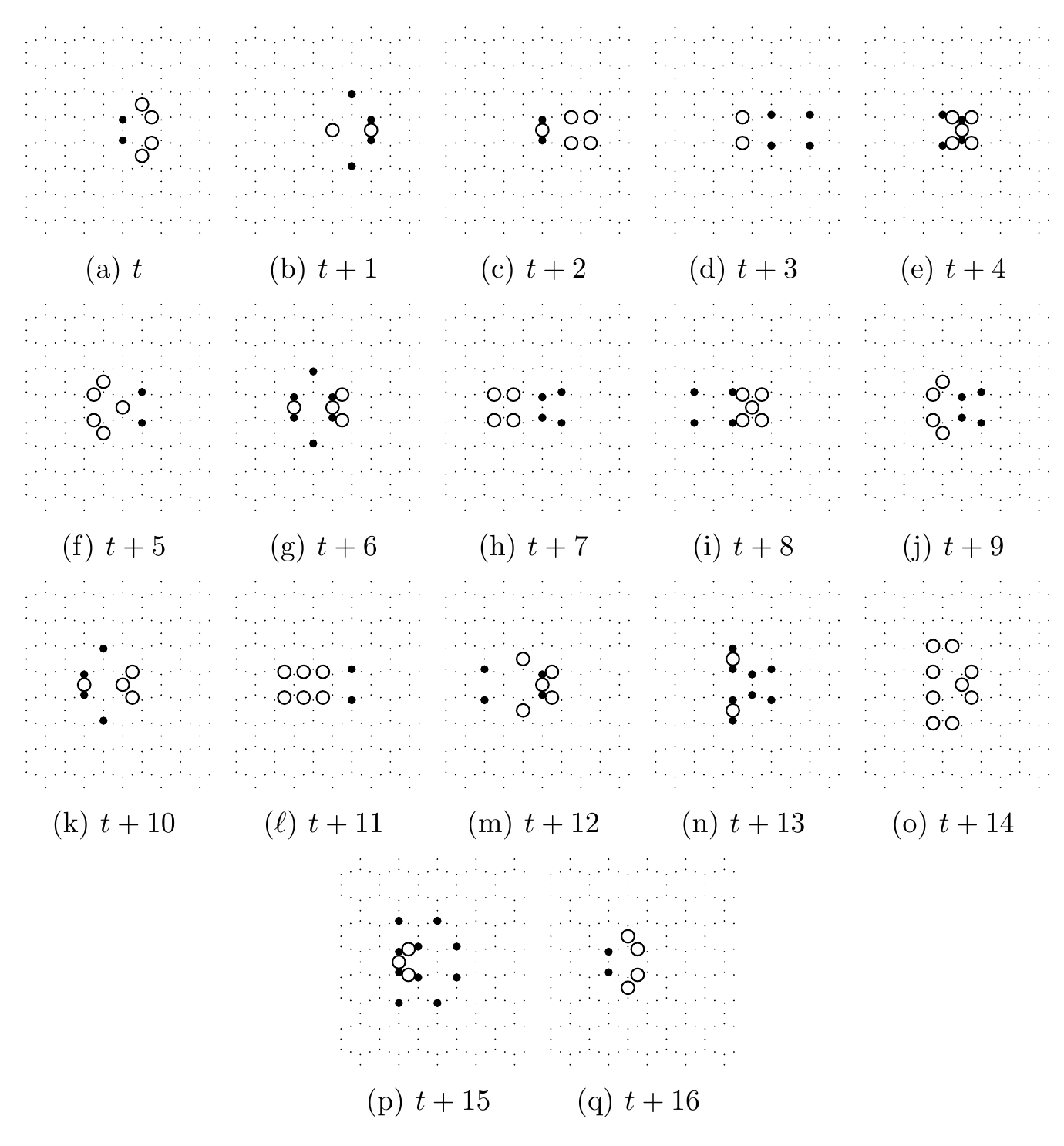}
\caption{Glider $G_{65}^{16}$ in rule $\mathcal{R}_{65}$. Glider moves west. 
Resting nodes are shown by dots. Non-resting silicon nodes are shown by small black discs. Non-resting oxygen nodes are shown by large 
circles.}
\label{5_2_16_65_gliderwest}
\end{figure}

\begin{figure}[!tbp] 
\centering
\includegraphics[scale=1]{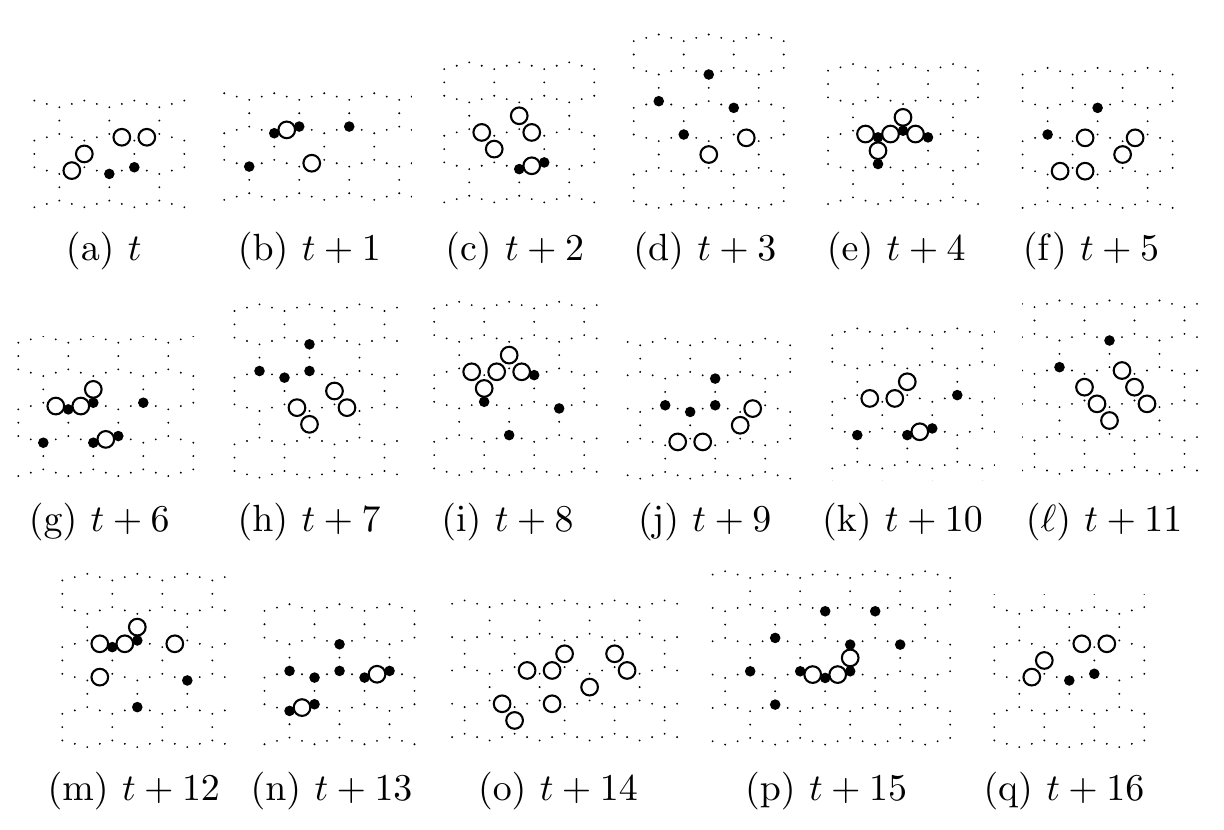}
\caption{Glider $G_{65}^{16}$ in rule $\mathcal{R}_{65}$. Glider moves south-east. Resting nodes are shown by dots. Non-resting silicon nodes are shown by small black discs. Non-resting oxygen nodes are shown by large 
circles.}
\label{5_2_16_65_glidersoutheast}
\end{figure}

\begin{figure}[!tbp] 
\centering
\includegraphics[scale=1]{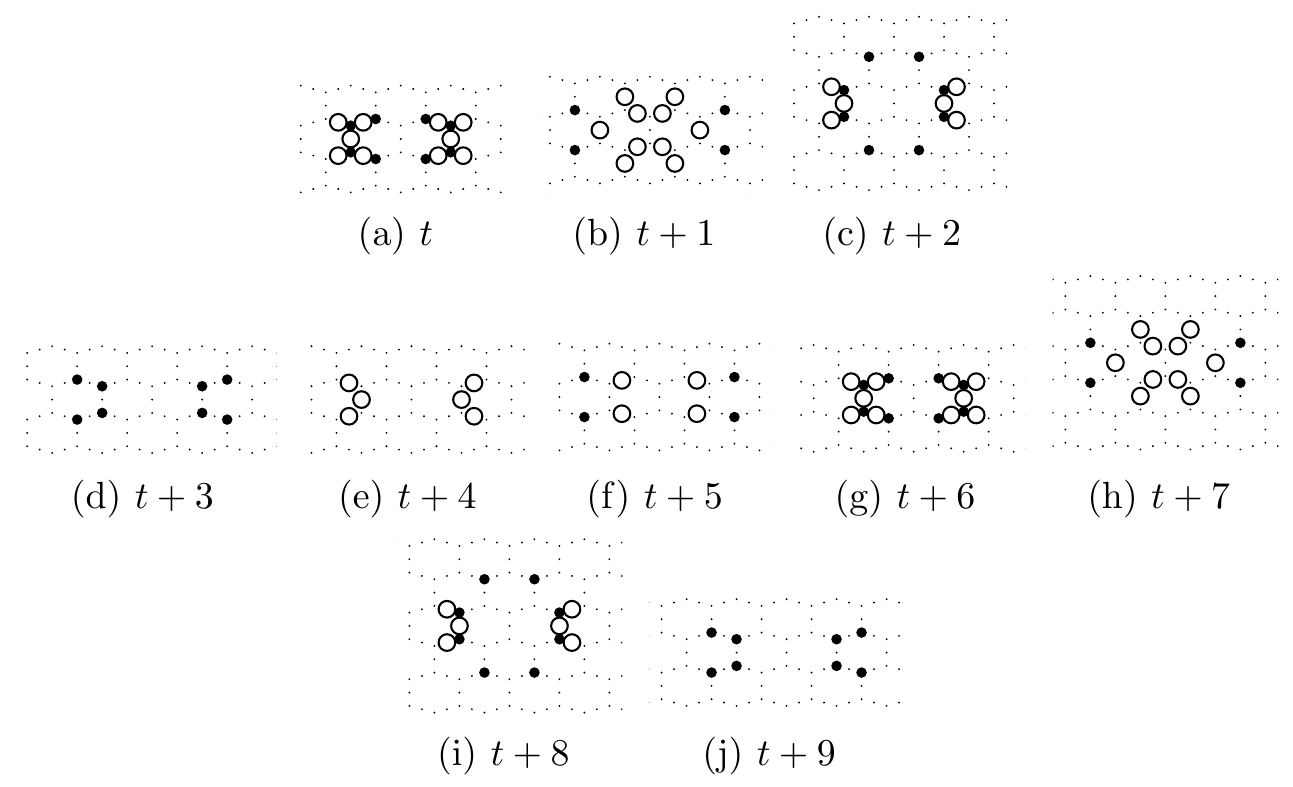}
\caption{Head on collision of gliders in rule $\mathcal{R}_{65}$, zero vertical shift. 
Glider travelling east collides with glider travelling west. Resting nodes are shown by dots. Non-resting silicon nodes are shown by small black discs. Non-resting oxygen nodes are shown by large 
circles. }
\label{5_2_16_65_glidercollision}
\end{figure}

Glider $G_{65}^{16}$ is observed in configurations produced by rule $\mathcal{R}_{65}$, see examples of gliders
travelling west and south-east in Figs.~\ref{5_2_16_65_gliderwest} and \ref{5_2_16_65_glidersoutheast}. Glider 
$G_{65}^{16}$ is theaviest amongst gliders discovered (Tab.~\ref{table}). It has longest period --- 16 time steps 
--- and maximum weight of 13 non-resting nodes (Figs.~\ref{5_2_16_65_gliderwest}p and \ref{5_2_16_65_glidersoutheast}p).  The glider undergoes  thickening and lengthening phases. In thickening phase 
the glider is larger, in terms of maximum between non-resting nodes, along a normal to its velocity vector, see e.g. 
Figs.~\ref{5_2_16_65_gliderwest}abfjko.  In lengthening phase glider is larger along parallel to its velocity vector, see 
e.g. Figs.~\ref{5_2_16_65_gliderwest}dhjl. Ratio of silicon to oxygen non-resting nodes changes during the glider's cycle as follows: $\frac{2}{4}$, $\frac{4}{2}$, $\frac{2}{5}$, $\frac{4}{2}$, $\frac{4}{5}$, $\frac{2}{5}$,
$\frac{6}{4}$, $\frac{4}{4}$, $\frac{4}{5}$, $\frac{4}{4}$, $\frac{4}{4}$, $\frac{4}{4}$, $\frac{2}{6}$,
$\frac{4}{5}$, $\frac{8}{2}$, $\frac{0}{9}$, $\frac{10}{3}$ and $\frac{2}{4}$.

Most collisions between gliders in rule $\mathcal{R}_{65}$ lead to explosions: an unlimitedly growing pattern is formed 
in the result of gliders interaction. Head on collision with nil vertical or horizontal shift between gliders is a rare case of collision where gliders do not explode (Fig.~\ref{5_2_16_65_glidercollision}) but fuse into an oscillator. One glider travels east, left non-resting pattern in Fig.~\ref{5_2_16_65_glidercollision}a, another glider travels west, right non-resting pattern in Fig.~\ref{5_2_16_65_glidercollision}a. The gliders come into `contact' with each other (Fig.~\ref{5_2_16_65_glidercollision}b) and form a transient pattern (Fig.~\ref{5_2_16_65_glidercollision}b) of 4 non-resting silicon nodes and 10 non-resting oxygen nodes. This pattern is then transformed into an oscillator of period six (Fig.~\ref{5_2_16_65_glidercollision}d--j).

\begin{figure}[!tbp] 
\centering
\includegraphics[scale=1]{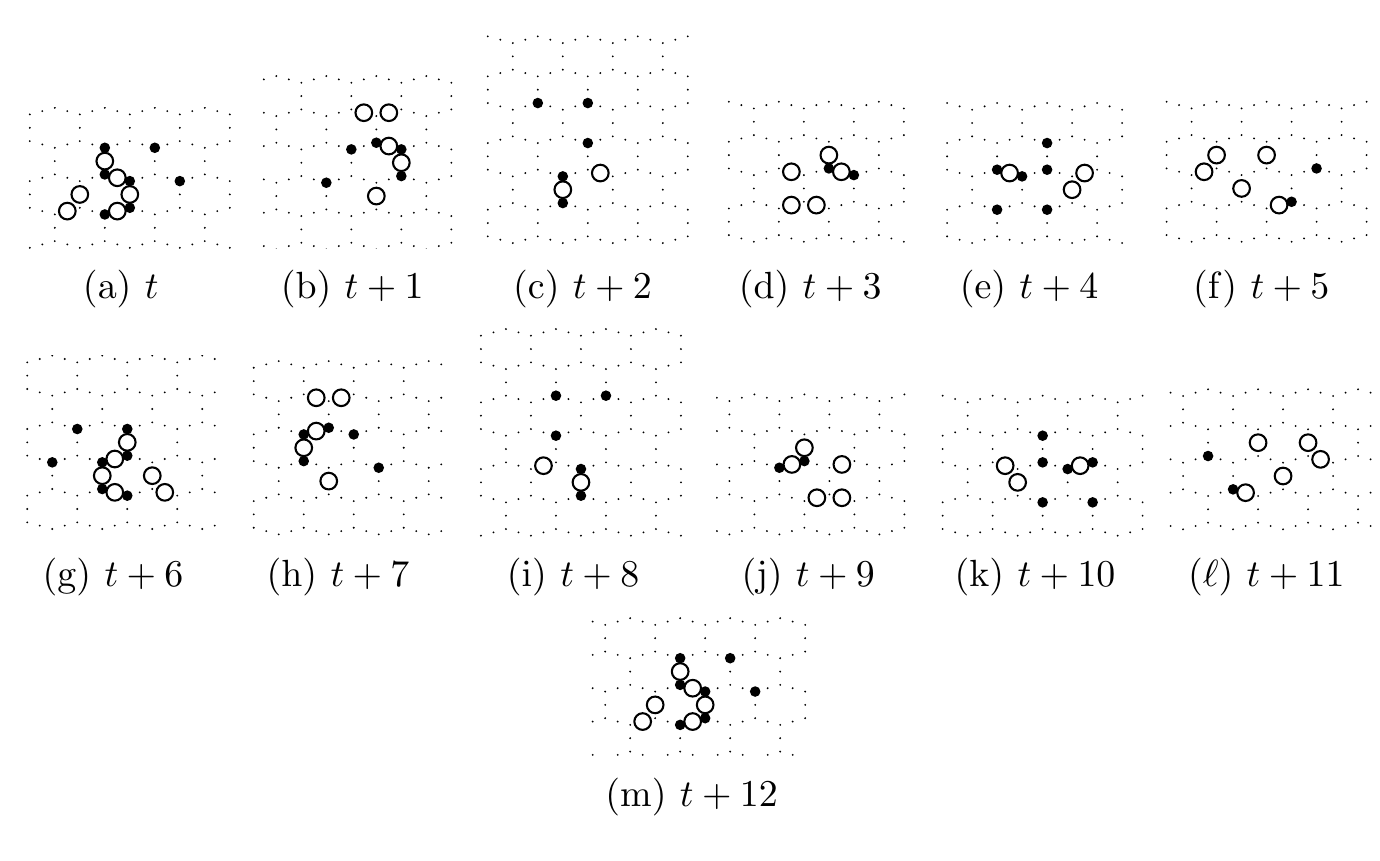}
\caption{Glider $G_{68}^{12}$ in  rule $\mathcal{R}_{68}$. 
Glider moves down from north straight to south. Resting nodes are shown by dots. Non-resting silicon nodes are shown by small black discs. Non-resting oxygen nodes are shown by large 
circles.}
\label{5_2_16_68_cheekyglider}
\end{figure}

\begin{figure}[!tbp] 
\centering
\includegraphics[scale=1]{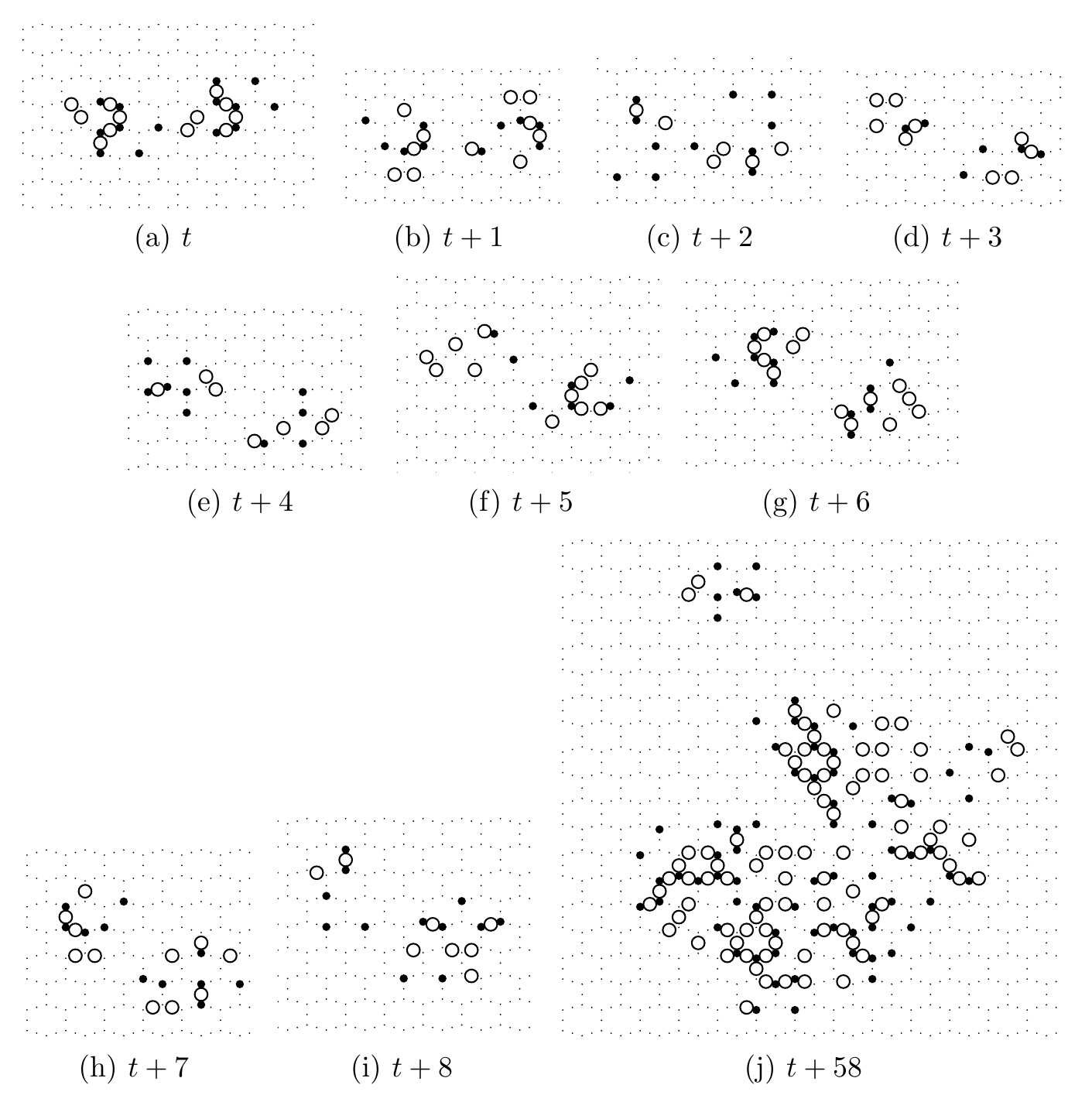}
\caption{Interaction of gliders $G_{68}^{12}$ in rule $\mathcal{R}_{68}$.  Glider moving north brushes with glider moving south. Resting nodes are shown by dots. Non-resting silicon nodes are shown by small black discs. Non-resting oxygen nodes are shown by large circles.}
\label{5_2_16_68_collisionOfglidersOneExplodes}
\end{figure}

Glider in rule $\mathcal{R}_{68}$ is shown in Fig.~\ref{5_2_16_68_cheekyglider}. The glider has period 12, and maximum weight 13 nodes (Tab.~\ref{table}). Glider reflects along its south-north axis during motion,  in middle of its cycle, namely glider state at time step $t$, $t=1, 2, \cdots$, is vertically symmetric to the glider state at time 
step $t+6$. The glider does not show a pronounced  thickening or lengthening but rather breathing around its centre of mass.  Ratios of non-resting silicon nodes to non-resting oxygen nodes in the glider's cycle are changed as follows: 
$\frac{7}{6}$,  $\frac{5}{5}$,  $\frac{5}{2}$,  $\frac{2}{5}$,  $\frac{6}{3}$,  $\frac{2}{5}$,  $\frac{7}{6}$,  
$\frac{5}{5}$,  $\frac{5}{2}$,  $\frac{2}{5}$,  $\frac{6}{3}$,  $\frac{2}{5}$,  $\frac{7}{6}$. 

Head on collision between two gliders in $\mathcal{R}_{68}$ without offset leads to explosion of both gliders. For certain offsets one glider can survive collision.  Such scenario is illustrated in 
Fig.~\ref{5_2_16_68_collisionOfglidersOneExplodes}. 
A glider moving north (left non-resting pattern in Fig.~\ref{5_2_16_68_collisionOfglidersOneExplodes}a) 
clips a glider moving south (right non-resting pattern in Fig.~\ref{5_2_16_68_collisionOfglidersOneExplodes}a).
The glider moving north does not suffer any damage and continues its journey undisturbed.  
The glider moving south becomes tangled (Fig.~\ref{5_2_16_68_collisionOfglidersOneExplodes}b--d) and 
explodes  (Fig.~\ref{5_2_16_68_collisionOfglidersOneExplodes}e--i and j).

\begin{figure}[!tbp] 
\centering
\includegraphics[scale=1]{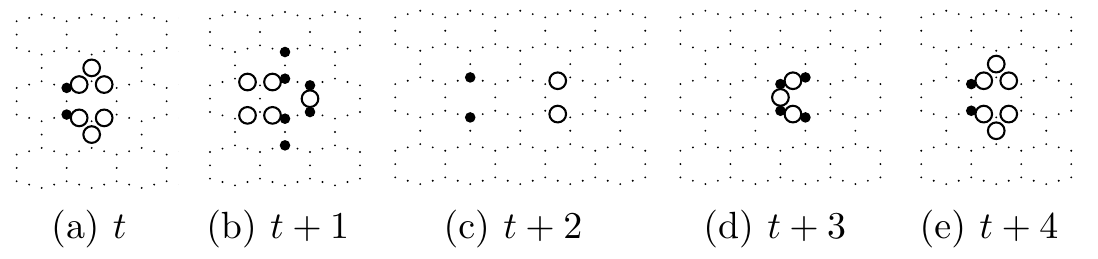}
\caption{Glider $G_{72}^{4}$ in rule $\mathcal{R}_{72}$. The glider travels east. Resting nodes are shown by dots. Non-resting silicon nodes are shown by small black discs. Non-resting oxygen nodes are shown by large 
circles. }
\label{5_2_16_72_glider1}
\end{figure}

Glider in rule $\mathcal{R}_{72}$ is shown in Fig.~\ref{5_2_16_72_glider1}. It is smallest amongst gliders discovered 
and has smallest period. It has just four different states. Ratios of non-resting silicons to non-resting oxygens in the glider's cycle are $\frac{2}{6}$,  $\frac{6}{5}$,  $\frac{2}{2}$ and $\frac{4}{3}$.

\begin{figure}[!tbp]  
\centering
\includegraphics[scale=1]{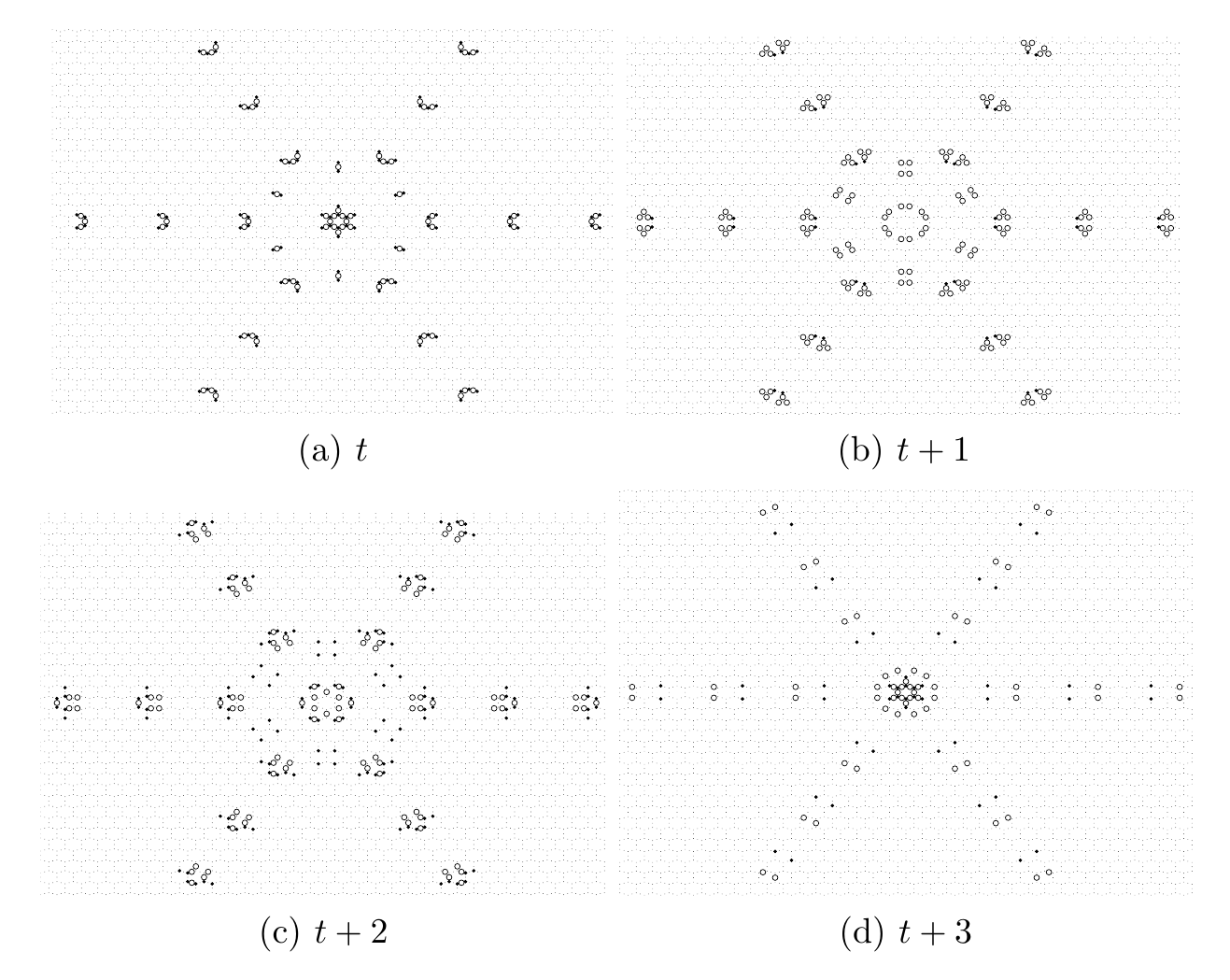}
\caption{Glider gun in rule $\mathcal{R}_{72}$.  Resting nodes are shown by dots. Non-resting silicon nodes are shown by small black discs. Non-resting oxygen nodes are shown by large 
circles.}
\label{glidergun}
\end{figure}

We did not find glider guns in rules $\mathcal{R}_{65}$ and $\mathcal{R}_{68}$, we believe they exist though and just wait to be found by someone patient. Glider gun in rule $\mathcal{R}_{72}$ is shown in Fig.~\ref{glidergun}.

\section{Oscillators}
\label{oscillators}

\begin{figure}[!tbp] 
\centering
\includegraphics[scale=1]{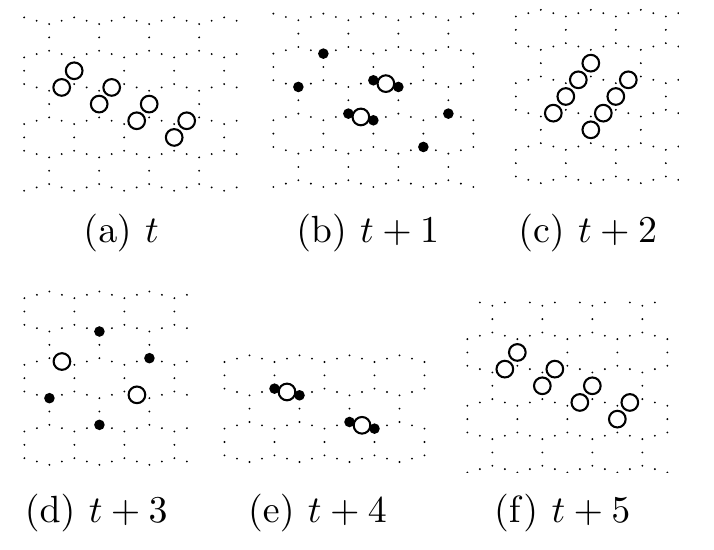}
\caption{Oscillator $O_{65}^5$ in rule $\mathcal{R}_{65}$. Resting nodes are shown by dots. Non-resting silicon nodes are shown by small black discs. Non-resting oxygen nodes are shown by large 
circles.}
\label{5_2_16_65_oscillator5}
\end{figure}

\begin{figure}[!tbp]  
\centering
\includegraphics[scale=1]{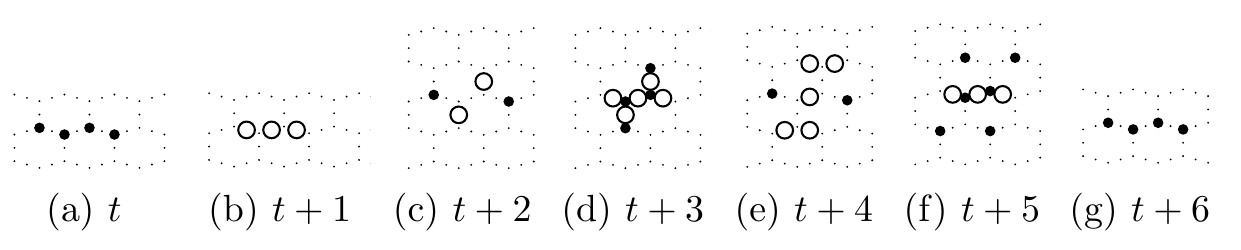}
\caption{Oscillator $O_{65}^6$ in rule $\mathcal{R}_{65}$. Resting nodes are shown by dots. Non-resting silicon nodes are shown by small black discs. Non-resting oxygen nodes are shown by large 
circles.}
\label{5_2_16_65_oscillator6}
\end{figure}

\begin{figure}[!tbp]   
\centering
\includegraphics[scale=1]{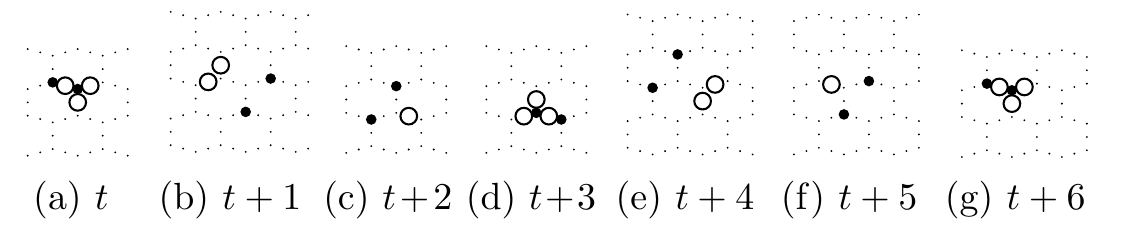}
\caption{Oscillator $O_{65}^{6A}$ in  rule $\mathcal{R}_{65}$. Resting nodes are shown by dots. Non-resting silicon nodes are shown by small black discs. Non-resting oxygen nodes are shown by large 
circles.}
\label{5_2_16_65_oscillator6B}
\end{figure}

Oscillator is a localised compact configuration of non-quiescent states which undergoes finite growth and 
modification but returns to its original state in a finite number of steps. Oscillators are generated by all three rules,
discussed in the paper, 
and can be easily found by randomly perturbing the automaton lattice.  Here we present a selection of most common oscillators. Some parameters of the oscillators are shown in Tab.~\ref{table}. Periods of oscillators vary from 2, 
oscillator $O_{72}^2$, to 12, oscillators  $O_{68}^{12}$ and $O_{72}^{12}$. Minimum weights vary from 2, 
oscillator $O_{72}^{12}$ to 6, oscillator $O_{65}^{5}$. Maximum weights vary from 4, oscillator
$O_{72}^{2}$, to 26, oscillator $O_{68}^{12}$. 

Three typical oscillators in rule  $\mathcal{R}_{65}$ are   shown in Fig.~\ref{5_2_16_65_oscillator5}, \ref{5_2_16_65_oscillator6} and \ref{5_2_16_65_oscillator6B}. Oscillator $O^5_{65}$ changes it configuration from 
two parallel chains of non-resting oxygen nodes oriented north-west to south-east (Fig.~\ref{5_2_16_65_oscillator5}a) 
and north-east to south-west  (Fig.~\ref{5_2_16_65_oscillator5}c) to sparse chains of non-resting silicons with few non-resting oxygens incorporated  (Fig.~\ref{5_2_16_65_oscillator5}bd). Oscillator $O^5_{65}$ is also oscillator in rule 
$\mathcal{R}_{72}$.

Oscillator $O^6_{65}$ switches its configuration between a single chain of non-resting silicons 
(Figs.~\ref{5_2_16_65_oscillator6}ag), a single chain of non-resting oxygens (Figs.~\ref{5_2_16_65_oscillator6}b), 
and regular non-linear arrangements of non-resting nodes  (Figs.~\ref{5_2_16_65_oscillator6}c--f). Oscillator 
 $O^{6A}_{65}$  is one of smallest oscillators (Fig.~\ref{5_2_16_65_oscillator6B}). It has period six. The oscillator configuration at time step $t$ is symmetric --- along north-east to south-west axis --- to its configuration at time step $t+3$.

\begin{figure}[!tbp]  
\centering
\includegraphics[scale=1]{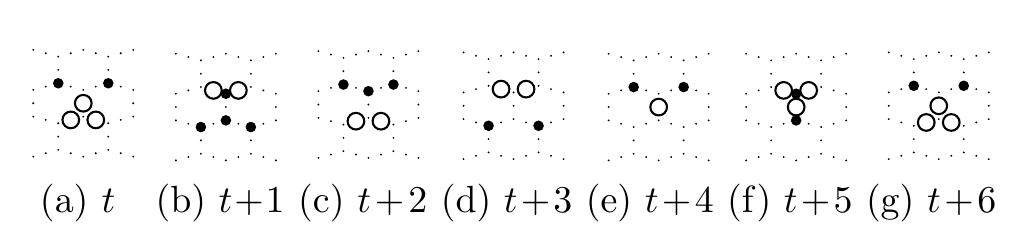}
\caption{Oscillator $O_{68}^6$ in rule $\mathcal{R}_{68}$. Resting nodes are shown by dots. Non-resting silicon nodes are shown by small black discs. Non-resting oxygen nodes are shown by large 
circles.}
\label{5_2_16_68_oscillator6}
\end{figure}

\begin{figure}[!tbp]  
\centering
\includegraphics[scale=1]{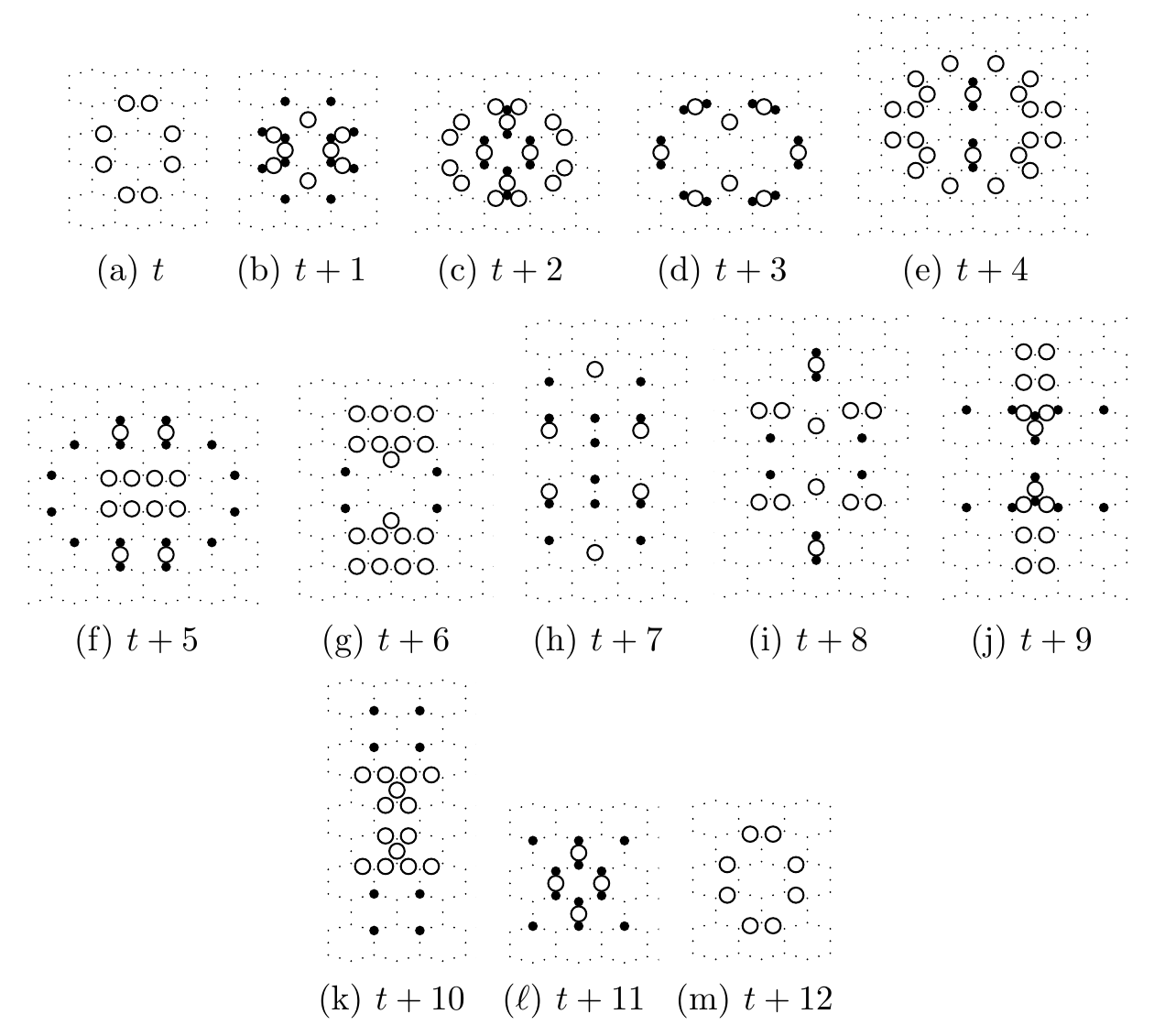}
\caption{Oscillator $O_{68}^{12}$ in rule $\mathcal{R}_{68}$. Resting nodes are shown by dots. Non-resting silicon nodes are shown by small black discs. Non-resting oxygen nodes are shown by large 
circles.}
\label{5_2_16_68_oscillator12}
\end{figure}

Exemplar oscillators discovered in rule  $\mathcal{R}_{68}$ are shown Figs.~\ref{5_2_16_68_oscillator6} 
and \ref{5_2_16_68_oscillator12}. Oscillator $O^6_{68}$ has the same configuration as oscillator 
$O^{6A}_{65}$, see  Fig.~\ref{5_2_16_68_oscillator6}f and Fig.~\ref{5_2_16_65_oscillator6B}adj.
Oscillator $O_{68}^{12}$ is the largest oscillator, amongst presented in the paper (Tab.~\ref{table}). 
The oscillator has period 16,  and consists of 28 non-resting nodes in its heaviest configuration 
(Fig.~\ref{5_2_16_68_oscillator12}f).

\begin{figure}[!tbp] 
\centering
\includegraphics[scale=1]{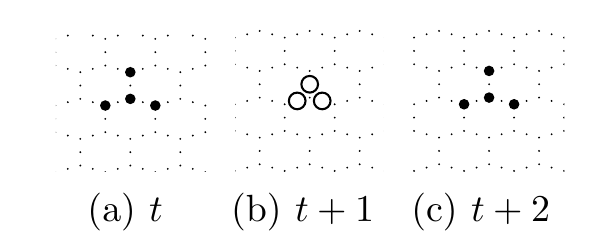}
\caption{Oscillator $O_{72}^2$ in rule $\mathcal{R}_{72}$. Resting nodes are shown by dots. Non-resting silicon nodes are shown by small black discs. Non-resting oxygen nodes are shown by large 
circles.}
\label{5_2_16_72_Oscillatorp3}
\end{figure}

Oscillator $O_{72}^2$ is the smallest one (Tab.~\ref{table}). It has just two configurations: one consists of four non-resting silicon nodes and another of three non-resting oxygen nodes (Fig.~\ref{5_2_16_72_Oscillatorp3}).

\begin{figure}[!tbp]  
\centering
\includegraphics[scale=1]{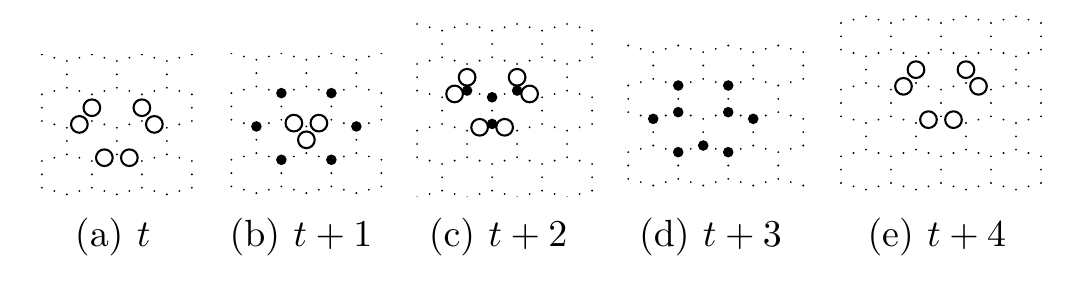}
\caption{Oscillator $O_{72}^4$ in rule $\mathcal{R}_{72}$. Resting nodes are shown by dots. Non-resting silicon nodes are shown by small black discs. Non-resting oxygen nodes are shown by large 
circles.}
\label{5_2_16_72_oscillatorp8}
\end{figure}

\begin{figure}[!tbp]  
\centering
\includegraphics[scale=1]{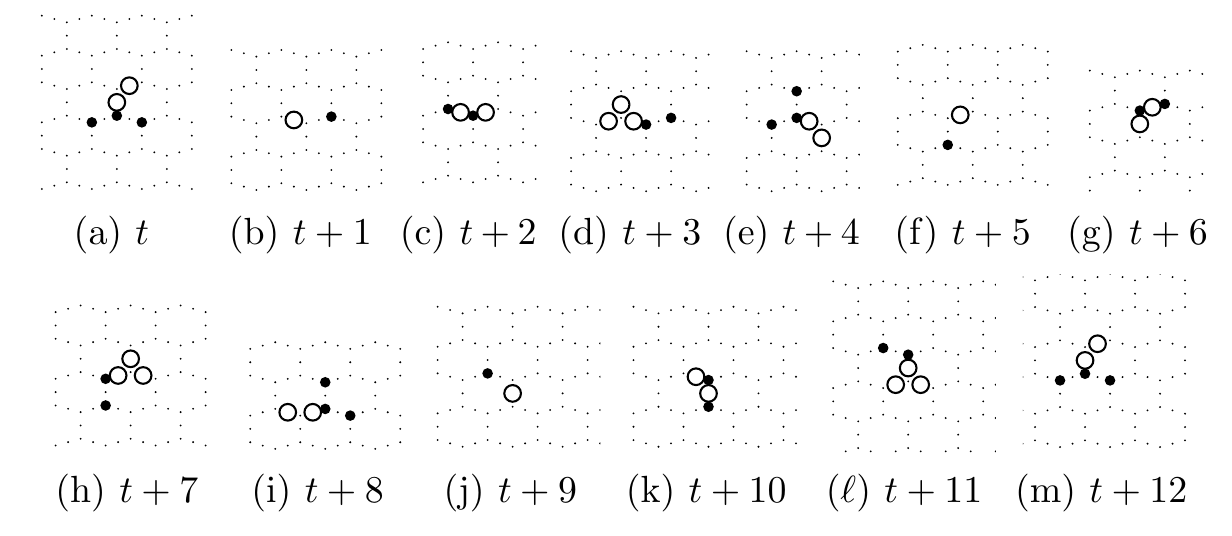}
\caption{Oscillator $O_{72}^{12}$ in rule $\mathcal{R}_{72}$. Resting nodes are shown by dots. Non-resting silicon nodes are shown by small black discs. Non-resting oxygen nodes are shown by large 
circles.}
\label{5_2_16_72_oscillatorp10}
\end{figure}

Oscillator  $O_{72}^{12}$ (Fig.~\ref{5_2_16_72_oscillatorp10}) is the smallest oscillator with longest period. Its period is 12 time steps and maximum weight is 5 non-resting nodes, see e.g. Fig.~\ref{5_2_16_72_oscillatorp10}d. 
When glider $G^4_{72}$ collides to $O_{72}^{12}$  there are three possible outcomes, depending on state of the oscillator at the collision and relative positions of the glider and the oscillator: oscillator and glider vanish in the result of collision, both localizations explode, and glider disappear but oscillator remains intact. In Conway's Game of Life an eater is a still life (stationary localisation that does not change its configuration) which repairs itself after collision with a glider or any other travelling localisation, while the colliding travelling localisation vanishes. Oscillator  $O_{72}^{12}$ is not a still life so it can be classified as oscillator-eater.

\section{Still life}
\label{stilllife}

\begin{figure}[!tbp]  
\centering
\includegraphics[scale=1]{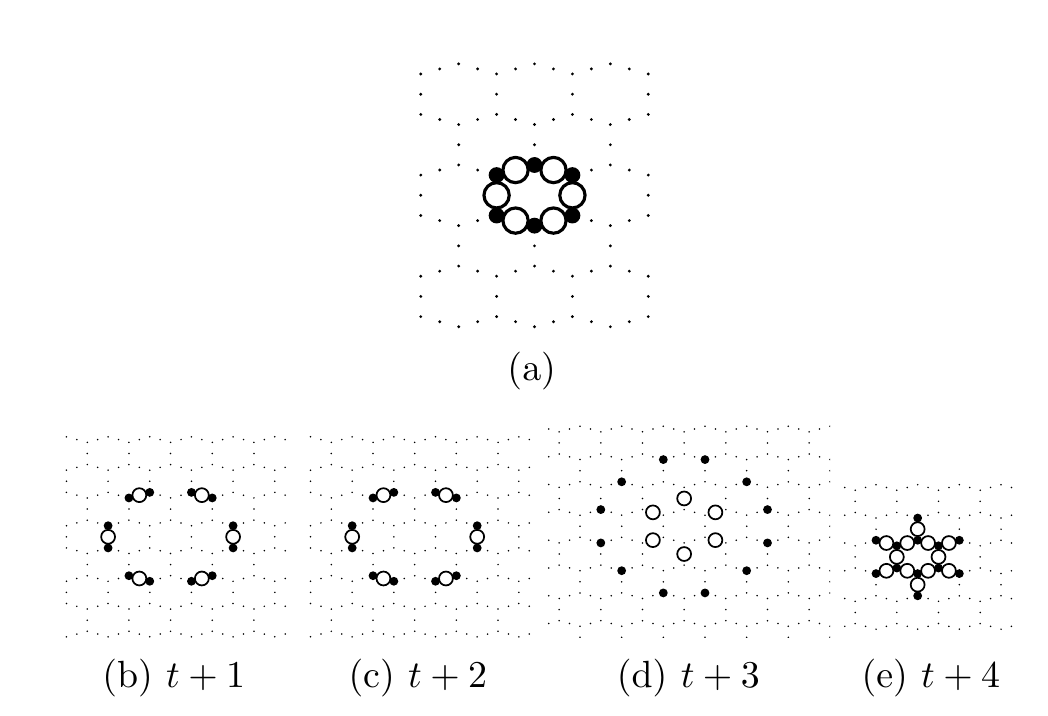}
\caption{Still localisation. (a)~Still localisation in rule $\mathcal{R}_{68}$; in this rule the localisation 
acts as eater. The localisation (a) becomes an oscillation (b--e) in rule $\mathcal{R}_{65}$. 
Resting nodes are shown by dots. Non-resting silicon nodes are shown by small black discs. Non-resting oxygen nodes are shown by large 
circles.}
\label{beehive}
\end{figure}

Still lifes are localizations that do not change their state/pattern or a position during automaton development. So far we found still lifes only in rule $\mathcal{R}_{68}$: this is a ring of  six oxygen and six silicon non-resting nodes as shown in Fig.~\ref{beehive}a. This patterns becomes an oscillator in rule $\mathcal{R}_{65}$ (Fig.~\ref{beehive}b) and is transformed into a glider gun in rule $\mathcal{R}_{68}$ (Fig.~\ref{glidergun}).

\section{Discussion}
\label{discussion}

\begin{figure} 
\centering
\includegraphics[width=0.9\textwidth]{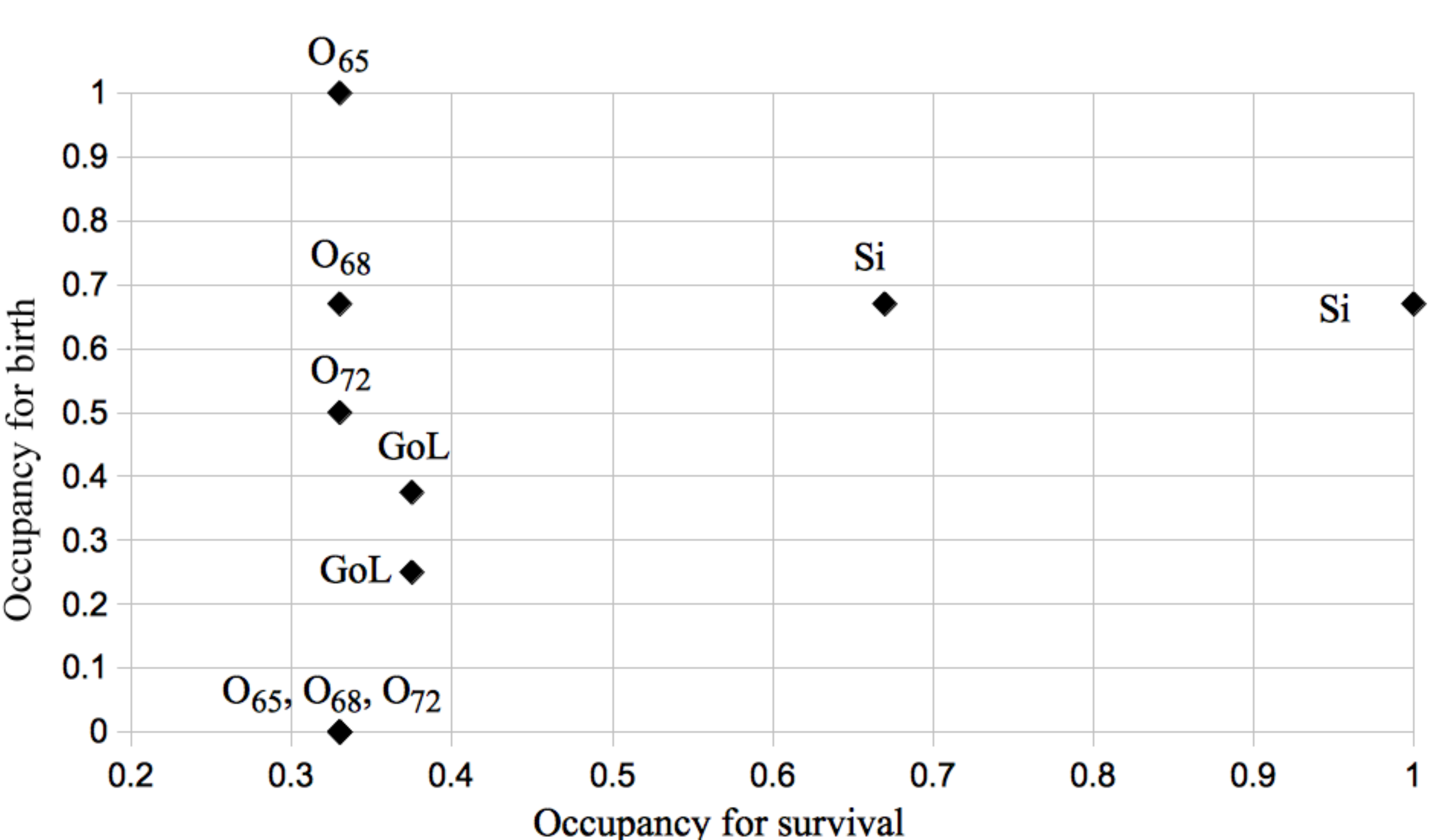}
\caption{Positions of Game of Life and phyllosilicate automata rules on `neighbourhood occupancy necessary for birth', transition from state 0 to state 1, versus  `neighbourhood occupancy necessary for birth', transition from state 1 to state 1.}
\label{occupancy}
\end{figure}

We introduced a regular automaton network which mimics topology of a silicate sheet and studied
node state transition rules which exhibit behaviour similar to Conway's Game of  Life automata.

Classical Conway's Game of Life automata  have rule B3/S23: a dead cell become alive, born if it has three live neighbours, a living cell survives if it has two or three live neighbours.  In notation of Life-like automata the rules can be written as follows. All silicon nodes update their states by rule  B23/S23.  Rules of oxygen nodes are B2/S06 (rule $\mathcal{R}_{65}$),  B2/S04 (rule $\mathcal{R}_{68}$) and  B2/S03 (rule $\mathcal{R}_{72}$).  Such notations could be somewhat misleading. In Conway's Game of Life a node (cell) has eight neighbours while in phyllosilicate automata a silicon node has three neighbours and oxygen node six neighbours. Thus, rules should be normalised by dividing number of non-resting neighbours necessary for birth and survival by neighbourhood size. Normalised rules are as follows. Normalised  Conway's Game of Life rule is B$\frac{3}{8}$/S$\frac{2}{8}$$\frac{3}{8}$. Rule for silicon nodes in phyllosilicate  is B$\frac{2}{3}$1/S$\frac{2}{3}$. Rules for oxygen nodes are B$\frac{2}{6}$/S01 ($\mathcal{R}_{65}$),  B$\frac{2}{6}$/S0$\frac{4}{6}$ ($\mathcal{R}_{68}$), and  B$\frac{2}{6}$/S0$\frac{3}{6}$ ($\mathcal{R}_{72}$).  On the map   `neighbourhood occupancy necessary for birth', transition from state 0 to state 1, versus  `neighbourhood occupancy necessary for birth'   (Fig.~\ref{occupancy}) Game of Life rules reside in the quadrant with below average 
occupancies. Occupancies necessary for silicon nodes to take non-resting state and to remain in the non-resting reside in the quadrant of above average occupancies.  Rules for oxygen nodes are characterised by below average occupancies necessary for survival  (Fig.~\ref{occupancy}). Distribution of rules on this map hints that conditions for survival are more critical than conditions for birth. Most rules supporting gliders obey a neigbourhood occupancy between 0.3 and 0.4 while neighbourhood occupancy for birth varies from 0 to 1. 

 We presented three types of node-state transition rules of a regular automaton network which mimics topology of a silicate sheet.  All three rules studied support travelling and oscillating stationary localizations. Yet only one rule exhibits still localizations. Our search for phyllosilicate automata exhibiting Conway's Game of Life behaviour was extensive yet not  exhaustive. Thus substantial chances remain that glider guns could be found in rules $\mathcal{R}_{65}$  and $\mathcal{R}_{68}$, and still lifes in rules $\mathcal{R}_{65}$  and $\mathcal{R}_{72}$.  
 
 Most interactions between localizations  lead to explosions of the localizations, when unlimitedly growing patterns of non-resting states are formed. Just few scenario of collisions lead to annihilation of both or at least one of colliding localizations. This may somehow limit, however not totally take away,  an applicability of phyllosilicate automata  in design of collision-based circuits~\cite{adamatzky_CBC}. Further studies in the interaction of localizations are required to make a definite conclusion about usefulness of phyllosilicate automata in the unconventional computing field. 

How gliders, oscillators and still lifes could be represented in real phyllosilicate lattices?

We believe they will be seen as domains of localised defects, e.g. recombination-induced defects formation~\cite{tanimura_1983} occurred in tetrahedra, and  vacancy-impurity pairs or isolated interstitial silicon defects~\cite{watkins_1977}. The imperfection, vacancy, and self-interstitial, substitutional and interstitial impurities are important because they are necessary for setting up travelling localizations of e.g. a substitutional impurity atom through the lattice. The travelling impurities cause local  deformations of the phyllosilicate lattice, e.g. via recombination of vacancies and carriers, these deformations produce waves. Thus, for example, state `1' of a node can represent a point interstitial defect, where the host atom moves to a non-lattice position~\cite{sokolski_1967}. The point defects can form travelling localizations, phenomenologically similar to gliders. Such localizations of defects may not be stable, even at low temperature, and disappear. However, duration of their life time could be enough to execute one or more collision-based logical gates. 

 A typical scenario of generating a glider on a phyllosilicate lattice would be the one detailed in~\cite{janavicus_2007}.
 A locally irradiation of the lattice with 1-3 MeV electrons, gamma rays, or fast neutrons~\cite{janavicus_2007} would 
 move silicon atoms into metastable interstitial states. Thus a negatively charged travelling vacancy is formed. Direction of 
 this localisation travelling could be controlled by inducing lattice vibrations and affecting electrons transitions. Obviously, to 
 obtain an experimental confirmation would be a very difficult task.

There is  at least some analogy between gliders-supporting automata rules and conditions for 
 emergence of a travelling vacancy. In all glider-supporting phyllosilicate automata studied a silicon node excites and stays excited if
 it has two or three excited neighbours. An oxygen is excited if it has two excited neighbours. As hinted in~\cite{janavicus_2007} negatively charged
 travelling vacancies can migrate only if at least two inter-atomic bonds are broken. 
 
When complemented by  molecular-dynamics method~\cite{law_2000} or reaction-diffusion computational approach~\cite{velichko_2002}  of simulating defects propagation, our approach can be useful in designing close to reality computing 
circuits based on travelling defects and clusters of defects.  This will be a topic of future studies to find what 
exact structure of defect, localisation or dislocation~\cite{bulatov_2001, pizzagalli_2011,sharma_1990} corresponds corresponds 
to gliders and still lifes.


\end{document}